\documentclass{wscpaperproc}
\usepackage{latexsym}
\usepackage{graphicx}
\usepackage{mathptmx}
\usepackage[T1]{fontenc}

\usepackage{amsmath}
\usepackage{amsfonts}
\usepackage{amssymb}
\usepackage{amsbsy}
\usepackage{amsthm}
\usepackage{booktabs}

\usepackage[pdftex,colorlinks=true,urlcolor=blue,citecolor=black,anchorcolor=black,linkcolor=black]{hyperref}

\newtheoremstyle{wsc}
{3pt}
{3pt}
{}
{}
{\bf}
{}
{.5em}
{}

\theoremstyle{wsc}

\begin{document}

%
%

\pagestyle{fancyplain}

\thispagestyle{plain}
\firstPageHead{}

\chead{\fancyplain{}{\itshape Zhang and Jiang}}

\rhead{}
\cfoot{}
\renewcommand{\headrulewidth}{0pt} 

\makeatletter
\let\@internalcite\cite
\def\cite{\def\@citeseppen{-1000}%
    \def\@cite##1##2{(##1\if@tempswa , ##2\fi)}%
    \def\citeauthoryear##1##2##3{##1 ##3}\@internalcite}
\def\citeNP{\def\@citeseppen{-1000}%
    \def\@cite##1##2{##1\if@tempswa , ##2\fi}%
    \def\citeauthoryear##1##2##3{##1 ##3}\@internalcite}
\def\citeN{\def\@citeseppen{-1000}%
    \def\@cite##1##2{##1\if@tempswa, ##2)\else{}\fi}%
    \def\citeauthoryear##1##2##3{##1 (##3)}\@citedata}
\def\citeA{\def\@citeseppen{-1000}%
    \def\@cite##1##2{(##1\if@tempswa , ##2\fi)}%
    \def\citeauthoryear##1##2##3{##1}\@internalcite}
\def\citeANP{\def\@citeseppen{-1000}%
    \def\@cite##1##2{##1\if@tempswa , ##2\fi}%
    \def\citeauthoryear##1##2##3{##1}\@internalcite}
\def\shortcite{\def\@citeseppen{-1000}%
    \def\@cite##1##2{(##1\if@tempswa , ##2\fi)}%
    \def\citeauthoryear##1##2##3{##2 ##3}\@internalcite}
\def\shortciteNP{\def\@citeseppen{-1000}%
    \def\@cite##1##2{##1\if@tempswa , ##2\fi}%
    \def\citeauthoryear##1##2##3{##2 ##3}\@internalcite}
\def\shortciteN{\def\@citeseppen{-1000}%
    \def\@cite##1##2{##1\if@tempswa, ##2\else{}\fi}%
    \def\citeauthoryear##1##2##3{##2 (##3)}\@citedata}
\def\shortciteA{\def\@citeseppen{-1000}%
    \def\@cite##1##2{(##1\if@tempswa , ##2\fi)}%
    \def\citeauthoryear##1##2##3{##2}\@internalcite}
\def\shortciteANP{\def\@citeseppen{-1000}%
    \def\@cite##1##2{##1\if@tempswa , ##2\fi}%
    \def\citeauthoryear##1##2##3{##2}\@internalcite}
\def\citeyear{\def\@citeseppen{-1000}%
    \def\@cite##1##2{(##1\if@tempswa , ##2\fi)}%
    \def\citeauthoryear##1##2##3{##3}\@citedata}
\def\citeyearNP{\def\@citeseppen{-1000}%
    \def\@cite##1##2{##1\if@tempswa , ##2\fi}%
    \def\citeauthoryear##1##2##3{##3}\@citedata}
%
%
%
\def\@citedata{%
    \@ifnextchar [{\@tempswatrue\@citedatax}%
                  {\@tempswafalse\@citedatax[]}%
}

\def\@citedatax[#1]#2{%
\if@filesw\immediate\write\@auxout{\string\citation{#2}}\fi%
  \def\@citea{}\@cite{\@for\@citeb:=#2\do%
    {\@citea\def\@citea{, }\@ifundefined
       {b@\@citeb}{{\bf ?}%
       \@warning{Citation `\@citeb' on page \thepage \space undefined}}%
{\csname b@\@citeb\endcsname}}}{#1}}%

%
\def\@citex[#1]#2{%
\if@filesw\immediate\write\@auxout{\string\citation{#2}}\fi%
  \def\@citea{}\@cite{\@for\@citeb:=#2\do%
    {\@citea\def\@citea{; }\@ifundefined
       {b@\@citeb}{{\bf ?}%
       \@warning{Citation `\@citeb' on page \thepage \space undefined}}%
{\csname b@\@citeb\endcsname}}}{#1}}%

%
\def\@biblabel#1{}
\makeatother



\newdimen\bibindent
\bibindent=0.0em
\def\thebibliography#1{\section*{\refname}\list
   {}{\settowidth\labelwidth{[#1]}
   \leftmargin\parindent
   \itemindent -\parindent
   \listparindent \itemindent
   \itemsep 0pt
   \parsep 0pt}
   \def\newblock{}
   \sloppy
   \sfcode`\.=1000\relax}


\setlength{\baselineskip}{12.7pt}

\title{A Large Language Model-Driven Agent-Based Modeling Framework with Multi-Round Communication for Simulating Vaccine Opinion Dynamics}

\author{\begin{center}Bo Zhang\textsuperscript{1,2}, Na Jiang\textsuperscript{1}\\ [11pt]
\textsuperscript{1}Thrust of Urban Governance and Design, The Hong Kong University of Science and Technology (Guangzhou) Guangzhou, Guangdong, CHINA\\
\textsuperscript{2}Jinhe Center for Economic Research, Xi'an Jiaotong University Xi'an, Shaanxi, CHINA\end{center}
}

\maketitle

\vspace{-12pt}

\section*{ABSTRACT}
Recently, Large Language Models (LLMs) have been utilized in various applications of computational social science and provide the possibility to integrate such models into agent-based modeling to explore the cognitive processes. However, how specific cognitive modules drive individual decisions and macro-level opinion dynamics remains unclear. Therefore, this study introduces a framework that integrates an LLM (Qwen3-8B) into agent-based modeling to investigate this problem, using vaccination opinion dynamics as a case study. We utilize this framework to simulate opinion dynamics among agents with heterogeneous profiles and social networks, evaluating scenarios by enabling different cognitive modules: a memory module and a prompt diversity module. The simulation results reveal that different cognitive modules have opposite impacts on our emergent opinion. Furthermore, the framework reproduces the non-linear behavior patterns of social influence observed in existing research, demonstrating our framework's validity and potential to reach the level 3 validation of agent-based models.






\section{INTRODUCTION}
\label{sec:intro}

Large Language Models (LLMs) have significantly advanced computational social science through applications in reasoning, zero-shot learning, and role-playing \shortcite{berti2025,zhang2023,zhou2024}. Among these applications, LLMs as computational proxies that mimic human behavior have transformed agent-based modeling from a static, rule-based heuristic method to a more complex approach \cite{horton2023}. Specifically, such an approach allows us to capture more realistic human decision-making by utilizing LLMs' generative capabilities for human-like behaviors, such as communication via dialogues and making decisions through these dialogues \shortcite{liu2025}. Consequently, agents can leverage an LLM to generate natural language to enrich their interaction, allowing researchers to simulate complex social systems and observe how macro-level phenomena emerge from micro-level language-based interactions \shortcite{larooij2025,lu2024,vanhee2025}.



With respect to integrating LLMs into agent-based modeling, one of the current efforts has been placed to extend core modules within the cognitive process by diversifying identity profiles, incorporating memory, reflection and react modules \shortcite{kuroki2025,park2023,shinn2023,yan2025}. Such modules enable agents to engage in complex dialogues and tasks \shortcite{yao2023} and are theoretically important for an individual's cognitive process \shortcite{chu2024,zhang2025}. However, it remains unclear which module from the individuals' cognitive processes contributes to their decision-making and further leads to the emergence of social phenomena on the macro-level  \shortcite{li2025,taillandier2026,zhou2026}. Therefore, the goal of this study is to explore how different cognitive modules shape macro-level social phenomena by simulating agents to engage in sustained and multi-round interactions driven by an LLM.

To achieve the goal above, opinion dynamics provides an ideal test environment for this study. Traditionally, simulations of opinion dynamics often treat peer influence as a simple transmission of states \cite{flache2017,yin2024}. However, real-world opinion formation, especially in high-stakes contexts like public health and vaccination, is rarely instantaneous; it relies heavily on multi-round, interactive dialogue and deliberation \shortcite{chuang2024,miranda2025}. However, one common point is that recent studies introducing LLMs to opinion dynamics often overlook these continuous conversational exchanges. Therefore, we would argue that an agent's ability to engage in sustained dialogue, memorize past dialogue, and reflect after the dialogue is essential for modeling realistic trajectories of consensus formation and opinion polarization. To study such opinion dynamics phenomena, this paper presents an LLM-driven agent-based modeling framework to simulate the opinion evolution related to vaccination opinion dynamics. 

In the rest of this paper, Section \ref{sec:methodology} provides the detailed design of this LLM-driven agent-based model, and then Section \ref{sec:results} demonstrates the simulation results of different scenarios. Finally, Section \ref{sec:conclusion} summarizes this work and identifies future works.




\section{METHODOLOGY}
\label{sec:methodology}

\subsection{Overview}
\label{subsec:overview}

\begin{figure*}[htbp]
\centering
\includegraphics[width=1\linewidth]{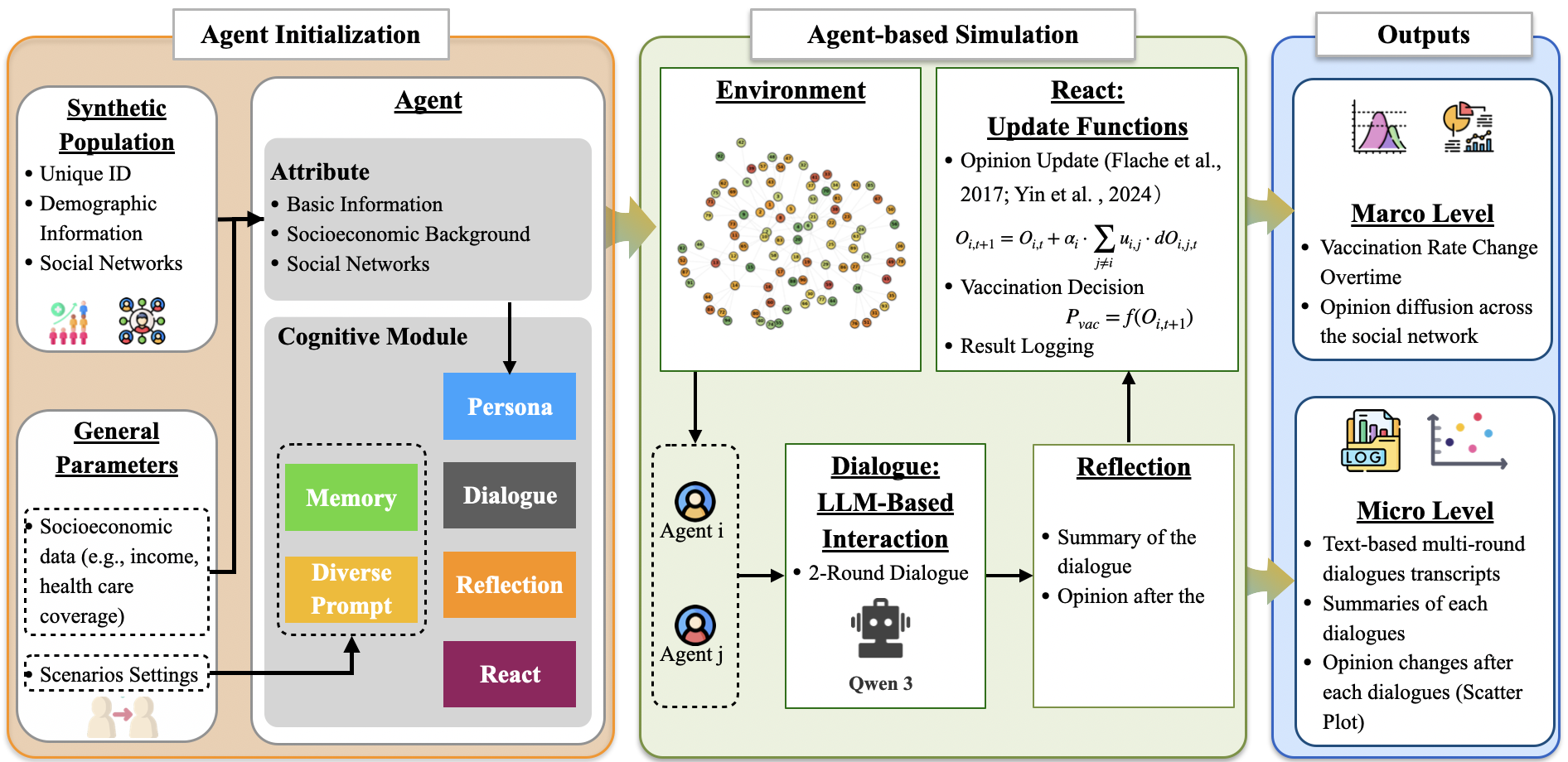} 
\caption{Overall framework of the LLM-enhanced agent-based simulation pipeline.}
\label{fig:framework}
\end{figure*}

This section details our framework for integrating LLMs into an agent-based model to simulate the changes in vaccination opinion dynamics. As shown in Figure~\ref{fig:framework}, the overall framework begins with initialization of agents along with their social networks by utilizing existing synthetic population and socioeconomic data. The core simulation engine generates multi-round dialogues between agents using the Qwen3-8B language model, selected for its strong performance and computational efficiency \shortcite{qwen3report}. Following each dialogue between two agents, agents will adjust their vaccination opinions according to the reflection on the dialogue. Then, the framework mathematically aggregates these opinion changes to update the final opinion after an agent's interactions with all neighbors connected by the social networks, which form discrete behavioral decisions and yield the macro-level outputs analyzed in Section~\ref{sec:results}.




\subsection{Agent and Cognitive Modules}
\label{subsec:initialization}

\subsubsection{Agent Initialization and Attributes}
To ground the simulation in realistic social structures, we initialize the agents using an existing synthetic population dataset built by \shortciteN{jiang2024large}. Recognizing that modern social influence stems from both physical and digital interactions \shortcite{ding2017asynchronous}, the environment incorporates the agents' associated household networks alongside a scale-free cyber network (i.e., social media network). From this dataset, we extract 30 agents from the same workplace along with 65 connected individuals from their extended networks (e.g., household and cyber), totaling 95 agents. We further enrich these agents with socioeconomic attributes such as education level and household income derived from Public Use Microdata Sample (PUMS) datasets \cite{acs2020pums}. This enrichment provides the critical demographic depth required to model nuanced vaccination decision-making. 

After initialization, the agents' attributes are summarized in Table \ref{tab:agent_attributes}. The demographic and socioeconomic attributes encompass critical attributes (e.g., age, gender, income, education level, and healthcare coverage). These attributes define the unique persona of each agent and provide the semantic foundation for the LLM to generate realistic dialogue. In addition, agents are connected by social networks, which serve as the environment driving the interactions. Specifically, different social networks dictate profile visibility during the interactions. Household network (\texttt{hh}) ties permit full profile disclosure between agents, while workplace network (\texttt{wk}) ties offer medium visibility by exposing a subset of socioeconomic data. Finally, social media network (\texttt{sm}) ties restrict visibility to basic traits like occupation and age. This profile visibility imitates the real-world information asymmetry among people with different demographic and socioeconomic backgrounds.


Other than demographic and socioeconomic attributes, each agent is assigned an initial opinion score $O_{i,0}$ indicating its opinion toward the vaccination, which is drawn from a normal distribution $\mathcal{N}(0, 0.3^2)$ and bounded within $[-1, 1]$. Additionally, an openness parameter ($\alpha$) controls the change of the opinion score in the next time step as shown in Equation \ref{eq:opinion_update}. It indicates that to what extent an agent's opinion could be impacted by the dialogue. In this model, $\alpha$ is uniformly set to 0.5 for all agents based on previous work \shortcite{yin2024}, which indicates that the agent's 50\% of opinion changes are caused by the dialogues, and the other 50\% is based on its initial opinion as discussed in Section \ref{subsec:outputs}. In addition, $Vaccination status$ represents whether the agent is vaccinated or not.


\begin{table}[ht]
\centering
\caption{Agent attributes and model parameters}
\label{tab:agent_attributes}
\renewcommand{\arraystretch}{1.2}
\small
\begin{tabular}{p{2.5cm} p{5.5cm} p{7.0cm}}
\toprule
\textbf{Category} & \textbf{Variables} & \textbf{Value Range} \\
\midrule
\textbf{Demographic \&} & Age & 18--77 (workplace members) \\
\textbf{Profile} & Gender & \{m, f\} \\
 & Urban status & \{0,1\} (rural/urban) \\
 & Personal income & \$0--\$104{,}020 \\
 & Health insurance coverage & \{has, no\} \\
 & Occupation category & 1--7 (mapped to text labels) \\
 & Education code/level & 11--24 (mapped to level text) \\
 & Family size & 1--5 \\
 & Opinion $O_{i,t}$ & $O_{i,t}\in[-1,1]$; at $t=0$: $O_{i,0}\sim\mathcal{N}(0,0.3^2)$, truncated to $[-1,1]$ \\
 & Openness ($\alpha$) & $0\le \alpha \le 1$; default $\alpha=0.5$ \\
 & Vaccination time step & 1--10 \\
 & Vaccination status& 0 or 1 (vaccinated or not) \\
\bottomrule
\end{tabular}
\end{table}

\subsubsection{Cognitive Modules}
As discussed in Section \ref{sec:intro}, this work aims to explore how different cognitive modules impact the overall changes in vaccination opinions. We designed six modules: persona-based foundation, dialogue, reflection, react, memory and diverse prompt modules. As for the persona-based module, it enriches agents' demographic profile and assigns an initial opinion score, $o_i \in [-1, 1]$, drawn from a normal distribution $\mathcal{N}(0, 0.3)$ using a fixed random seed. The dialogue module executes multi-round conversations. In the reflection module, the language model is utilized to review and summarize the multi-round dialogue and output an updated opinion score. The react module then incorporates the updated opinion scores from the reflection modules into the mathematical update functions shown in Figure~\ref{fig:framework}. This closes the loop: conversation leads to opinion change, which in turn leads to changes in vaccination status. The memory module enables agents to retain histories of their past opinions and dialogues. As the agents are enriched with socioeconomic attributes and connected with different social networks, their communication styles could diversify based on who the agent is and who the agent communicates with; the diverse prompt module introduces such variations to the agents.

\subsection{Agent-based Simulation: Main Process}
\label{subsec:main_function}

The central component of the framework is the Main Function shown in Figure \ref{fig:framework}, a cyclic process that dictates agent behavior and social influence at each time step. Two agents connected by the social network are selected to interact, initiating a two-round dialogue related to vaccination followed by a reflection process.

\begin{figure}[htbp]
\centering
\includegraphics[width=0.5\linewidth]{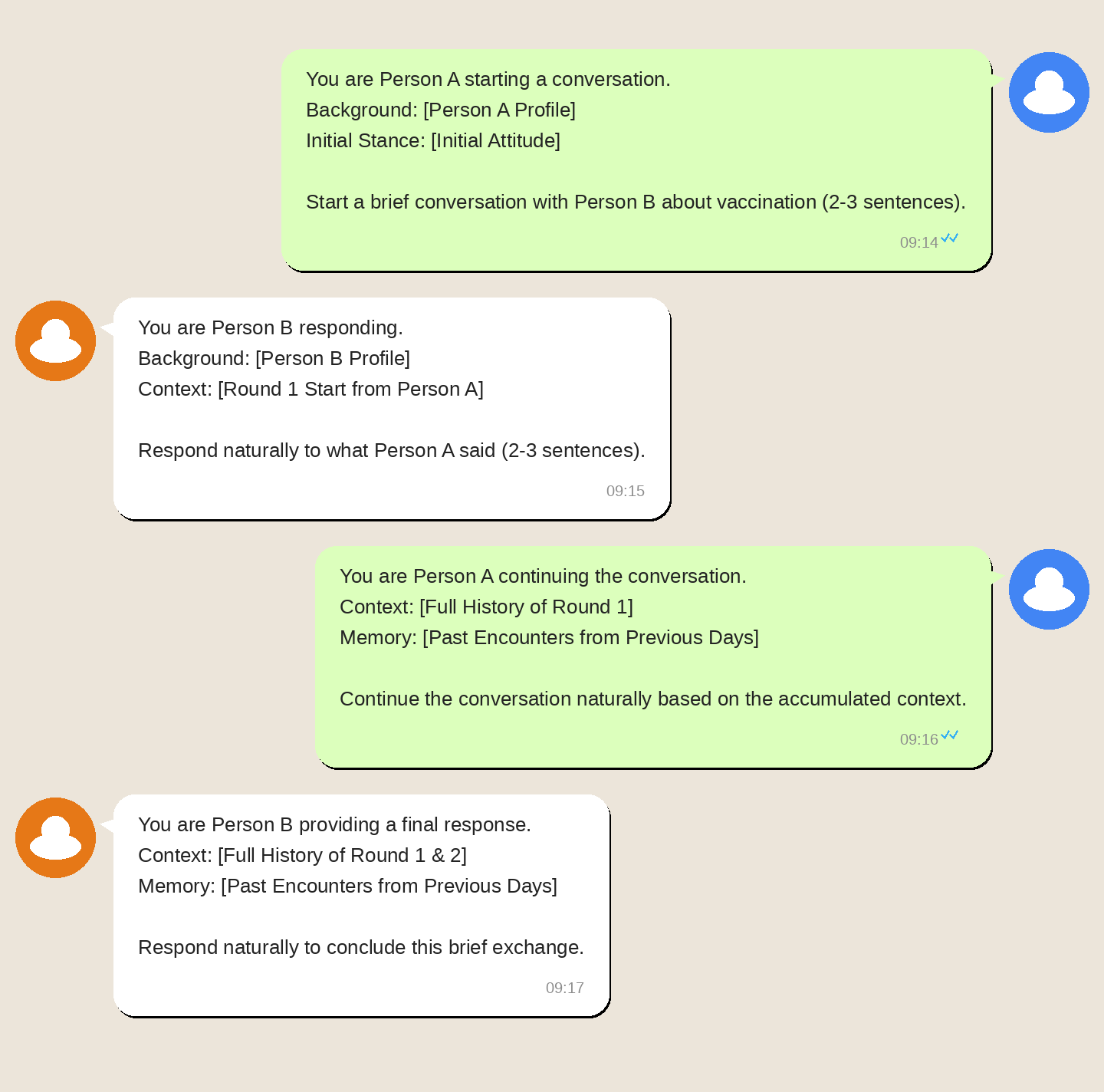}
\caption{Example of a multi-turn, realistic dialogue generated between two cognitive social agents.}
\label{fig:chat}
\end{figure}

These two-round dialogues are driven by the Qwen3-8B language model, which can be exemplified by the chat in Figure \ref{fig:chat}. The two-round dialogue starts with system prompt to ask Agent A to have a conversation with Agent B about vaccination, which utilized the Agents A's complete persona and current opinion score. Then Agent B will respond with the prompt to continue the dialogue using its own persona and vaccine opinion score.


Following the two-round dialogues, Agent B undergoes a reflection process driven by the reflection module (see Section \ref{subsec:initialization}). Within this module, the agent utilizes the Qwen3-8B language model to review the past dialogues and output one text-based summary and one numeric opinion score representing the agent's current opinion on vaccination.

The opinion score from the reflection module is integrated into the react module, which is the mathematical update functions to determine agent opinion and subsequent vaccination decisions. Drawing upon established social influence models \shortcite{flache2017,yin2024}, the opinion state of agent $i$ is updated as follows:

\begin{equation}
O_{i,t+1} = O_{i,t} + \alpha_i \cdot \sum_{j \neq i} u_{i,j} \cdot dO_{i,j,t}
\label{eq:opinion_update}
\end{equation}

where the new opinion $O_{i,t+1}$ incorporates the agent's internal openness parameter $\alpha_i$ and the weighted sum of social influences from all distinct neighbors $j$. The weighted influence $u_{i,j}$ is derived from the relational tie strength and visibility rules described in Section \ref{subsec:initialization}, and $dO_{i,j,t}$ represents the direct opinion shift derived from the reflection score following the dialogue.

Updated opinions inform the discrete vaccination decision process, modeled as $P_{vac} = f(O_{i,t+1})$. Agents whose updated opinions cross the threshold of $o_i > 0$ will have the probability (i.e., $P_{vac}$) to be vaccinated, representing a behavioral commitment to take vaccination. Once an agent is vaccinated, its opinion to vaccination will remain 1, and the vaccination status will change to 1.

\subsection{Expected Outputs}
\label{subsec:outputs}

By using this agent-based modeling framework, the simulation generates results from different levels related to vaccination opinion dynamics as shown in Figure \ref{fig:framework}. At the macro level, this framework continuously tracks the aggregated vaccination rate reflecting the agent's vaccination opinions or decisions, and visualizes opinion diffusion across the social networks. At the micro level, the framework captures the pairwise (i.e., two agents) opinion changes after each dialogue. To ensure the interpretability of the generative process and support rigorous qualitative scrutiny, the framework also compiles detailed interaction log files. These logs capture the full transcripts of all multi-round dialogues alongside the agents' summaries generated by the reflection module.

\section{SIMULATION RESULTS}
\label{sec:results}

Before demonstrating the final simulation results, we first conduct a verification process to ensure the simulation framework is working as designed (e.g., via code walkthroughs and visual debugging). Additionally, we ensure that the Qwen3-8B produces valid outputs. Since the model may occasionally refuse the role-play task due to safety alignment, generating responses containing phrases such as ``As an AI'' or ``I cannot provide,'' these cases are identified using a predefined keyword list and removed. The number of filtered responses is recorded at each step to monitor output stability. Overall, these checks confirm that this agent-based simulation framework is verified for scenario testing.

As this paper aims to explore the overall vaccination opinion dynamics through multi-round dialogue among agents with heterogeneous demographic and socioeconomic backgrounds (discussed in Section \ref{sec:intro}), we incorporate persona-based foundations together with dialogue, react, and reflection modules. The rationale for including a reflection module as part of the foundation is that the cognitive process is complex and the change in opinion is not instant, which need a reflection process \cite{KennedyModellingHumanBehaviour2012}. Thus, as Figure \ref{fig:framework} shows, the reflection module is enabled in all scenarios. 

By enabling memory and diverse prompt modules, we construct the four experimental scenarios summarized in Table \ref{tab:final_metrics}: 1) The baseline scenario relies only on the foundational components. 2) The memory scenario activates the memory module based on foundational modules. 3) The prompt diversity scenario enables the diverse prompt module to show the effect of communication styles. Finally, 4) the combined scenario integrates the entire modules to simulate the overall effects of two modules (i.e., memory and diverse prompt module). 

For the simulation details, we conduct 10 independent batch runs for each scenario above, and each run has 10 time steps to simulate 95 agents' dialogues via their social network. To make the outcomes of different scenarios comparable, we set identical initial opinion assignments for each scenario, and then we report the average results across 10 independent batch runs for each scenario, summarized in 
Table~\ref{tab:final_metrics}.

As shown in Table~\ref{tab:final_metrics}, a few patterns are visible. First, the prompt diversity scenario reaches the highest average opinion and vaccination rate, suggesting strong effects are brought by the dialogues. And the memory scenario consistently shows lower adoption outcomes, indicating resistance to opinion change. Second, the combined scenario does not simply average these effects, but instead exhibits higher polarization alongside moderate vaccination rate. Finally, the influence coefficient $r$ declines when additional modules are introduced. By comparing the outcomes from different scenarios controlled by different cognitive modules, the framework captures the preliminary emergence of various vaccination rates, final opinion distributions along with the degrees of opinion polarization on the macro level, but further analysis needs to be conducted for each scenario on both macro and micro levels.



\begin{table}[htb]
\centering
\caption{Final-step simulation statistics \label{tab:final_metrics}}
\resizebox{\columnwidth}{!}{%
\begin{tabular}{lccccc}
\toprule
\textbf{Configuration} & \textbf{Enabled Modules} & \textbf{Avg.\ Opinion} & \textbf{Polarization} & \textbf{Vax.\ Rate} & \textbf{Influence $r$} \\
\midrule
Baseline & None & $0.389 \pm 0.051$ & $0.596 \pm 0.010$ & $47.1\% \pm 4.8\%$ & $0.469$ \\
Memory & Memory Only & $0.184 \pm 0.053$ & $0.594 \pm 0.018$ & $32.9\% \pm 3.5\%$ & $0.386$ \\
Prompt Diversity & Diversity Only & $0.462 \pm 0.033$ & $0.590 \pm 0.012$ & $52.9\% \pm 3.2\%$ & $0.412$ \\
Combined & Memory + Diversity & $0.293 \pm 0.069$ & $0.606 \pm 0.016$ & $40.9\% \pm 6.1\%$ & $0.341$ \\
\bottomrule
\end{tabular}
}
\end{table}

\subsection{Macro-Level Dynamics: Vaccination Rates, Opinion Trajectories, and Polarization}
\label{subsec:macro_validation}

\begin{figure}[htbp]
    \centering
    \includegraphics[width=\linewidth]{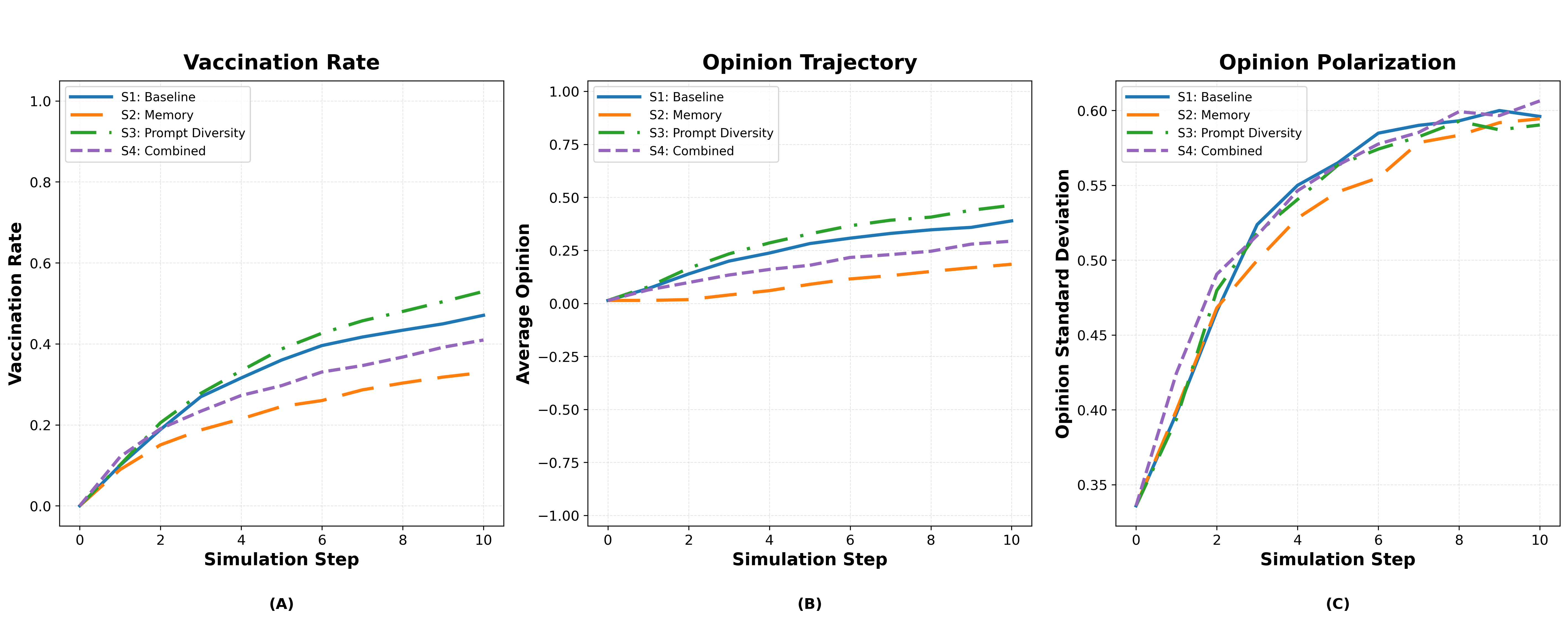}
    \caption{Comparison of Vaccination Rate and Opinion Dynamics of Different Scenarios}
    \label{fig:vax_opinion_comparison}
\end{figure}

Firstly, we further examine macro-level outcomes through the vaccination rate (Figure~\ref{fig:vax_opinion_comparison}A) and the evolution of average opinions and their polarization (Figure~\ref{fig:vax_opinion_comparison}B and C), which together reflect how micro-level interactions accumulate into the changes of collective patterns over time.

With respect to the overall vaccination rate (Figure~\ref{fig:vax_opinion_comparison}A), all four scenarios show an upward trend. The baseline scenario serves as a natural reference point, where agents interact under a relatively homogeneous rule, leading to a gradual and stable increase in both the average opinion and vaccination rate. However, introducing the prompt diversity module noticeably reshapes the overall trajectory, as agents are able to adopt different communication styles, which in turn leads to a substantially higher level of vaccination rate (see Figure~\ref{fig:vax_opinion_comparison}A). This pattern suggests that communication heterogeneity enhances communication effectiveness by utilizing different communication styles to start the dialogue based on their demographic and socioeconomic backgrounds. By contrast, enabling the memory module shifts the overall vaccination rate in the opposite trend and shows the lowest vaccination rate among the four scenarios (see Figure~\ref{fig:vax_opinion_comparison}A). This is because agents become less responsive to new interactions or dialogues, as the agent will remember the memory of the past dialogues and have the new dialogue with these memories. As for the combined scenario, the vaccination rate could be explained and averaged by these two opposite effects, which is relatively lower than the baseline vaccination rate.

As for the overall opinion score trend shown by Figure~\ref{fig:vax_opinion_comparison}B, similar to the vaccination rate, it also shows an upward trend, because the framework vaccination decision is positively correlated with the opinion (see Section \ref{sec:methodology}). This shows the effectiveness of the prompt diversity module and the resistance of the memory module as discussed above. Alongside this upward trend in average opinions, the framework successfully captures the emergence of opinion polarization. As shown in Figure~\ref{fig:vax_opinion_comparison}C, the upward trend is captured in the standard deviation of opinion, which means the opinion polarization emerges under all four scenarios. This is not only because most of the dialogues have positive impacts among agents, but also because the framework sets the opinion to 1 once an agent gets vaccinated.

Through measuring all four scenarios, the results of vaccination rate and opinion suggest that the prompt diversity module alone increases both of them. However, the combined scenario shows a lower vaccination rate and opinion score compared to the baseline, because of the resistance effect generated by the memory module, which balances out the effect brought by the prompt diversity module.

\subsection{Opinion Distribution Evolution and Analysis}
\label{subsec:distributional_nuance}

\begin{figure}[htbp]
    \centering
    \includegraphics[width=0.8\linewidth]{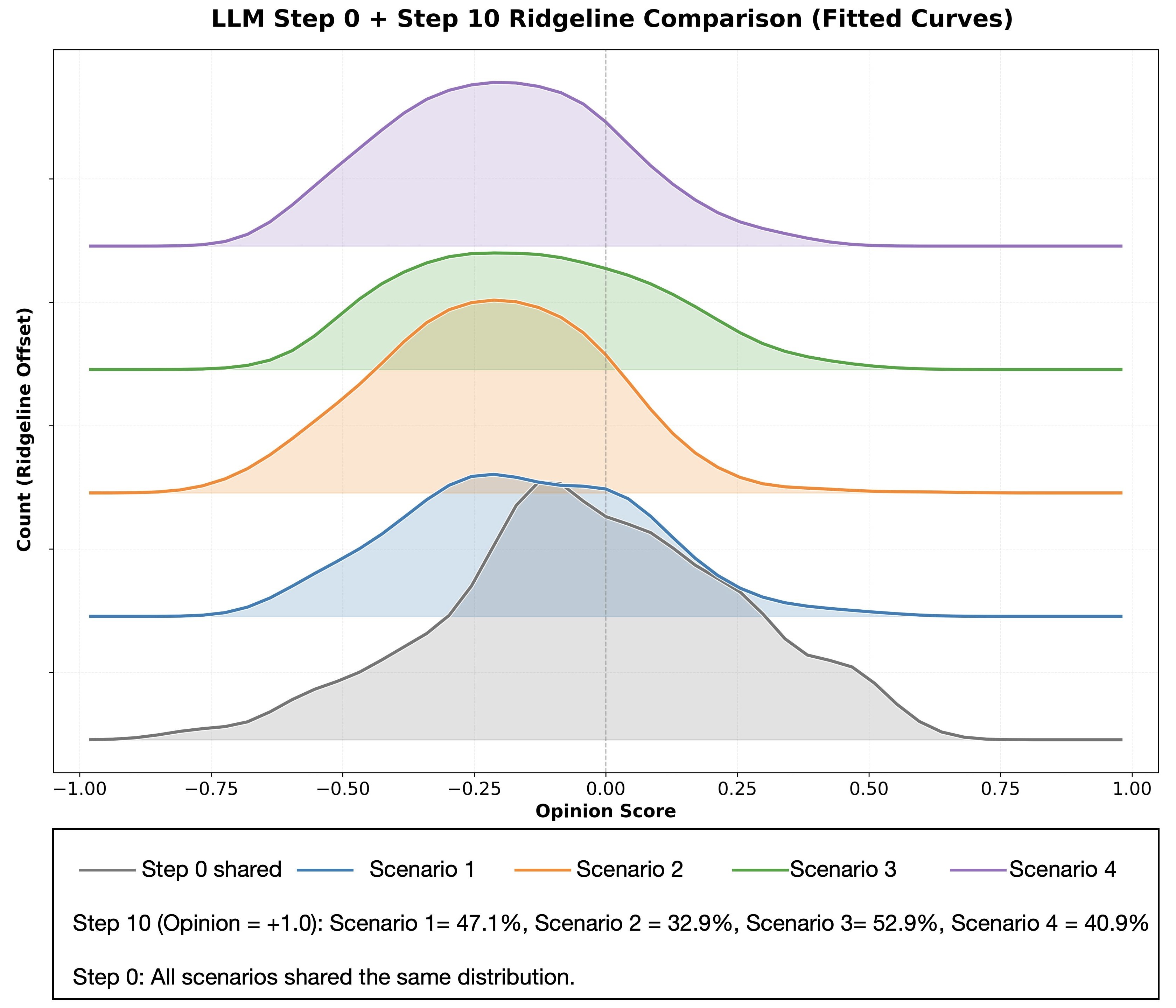}
    \caption{Ridgeline plot comparing opinion distributions from Step 0 to Step 10}
    \label{fig:opinion_distribution}
\end{figure}

Beyond average trends, it is also important to examine how the full distribution of opinions evolves over time. Figure~\ref{fig:opinion_distribution} presents a ridgeline plot comparing the initial distribution (Step 0) with the final distributions across different scenarios, allowing us to directly observe how opinions shift and spread. As discussed in Section \ref{sec:methodology}, agents who have successfully vaccinated change the opinion score to 1 are excluded in Figure~\ref{fig:opinion_distribution}. For the agents who are successfully vaccinated, the prompt diversity scenario produces the largest share of such agents ($52.9\%$), while the memory scenario yields the lowest ($32.9\%$) (see Figure~\ref{fig:opinion_distribution}).

Disregarding the vaccinated agents, each curve shows the final opinion distribution of all four scenarios and reveals how opinions evolve for agents who are not vaccinated. Compared to the baseline (i.e., blue curve in Figure~\ref{fig:opinion_distribution}), the prompt diversity scenario (i.e., green curve in Figure~\ref{fig:opinion_distribution}) shifts the distribution to the right, indicating that even when dialogue does not lead to vaccination, it still has positive impacts on agents' opinions. However, the memory scenario (i.e., orange curve in Figure~\ref{fig:opinion_distribution}) concentrates the probability mass on the negative side, suggesting that retaining interaction history reinforces the negative opinion of the vaccination. The combined scenario (i.e, purple curve in Figure~\ref{fig:opinion_distribution}) also shows a wide spread, with more mass appearing in both the positive and negative extremes. This indicates that some agents are impacted to get vaccinated, while others become more resistant, leading to a more polarized opinion distribution.

Overall, these distributional patterns complement the earlier results. Rather than shifting opinions uniformly, different modules reshape the distribution in distinct ways, with the memory module reinforcing negative attitudes and the combined scenario increasing opinion polarization. This provides additional evidence that the framework captures how heterogeneous responses at the individual level translate into polarization at the aggregate level.

\subsection{Micro-Level Influence Outcome: Scatter and Quadrant Analysis}
\label{subsec:micro_validation}

\begin{figure}[htbp]
    \centering
    \includegraphics[width=\linewidth]{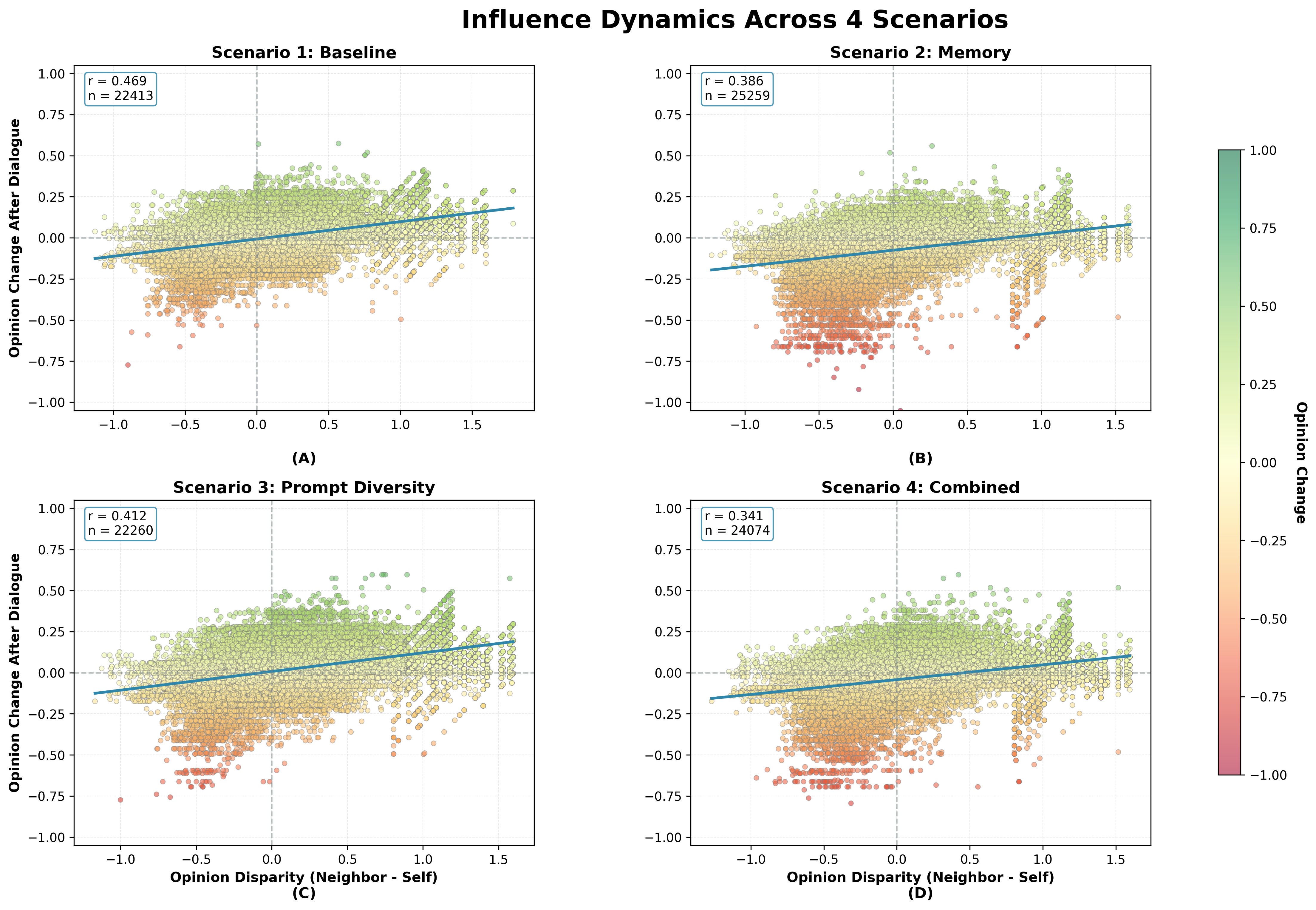}
    \caption{Influence Scatter Plots Across Four Scenarios}
    \label{fig:Scatter Plot}
\end{figure}

To demonstrate the effectiveness of our framework, we move from aggregate outcomes to the micro-level interaction process. Traditional opinion dynamics models typically assume a linear relationship: opinion changes are proportional to opinion distance. However, this assumption overlooks the cognitive modules. Figure~\ref{fig:Scatter Plot} directly examines this by plotting opinion disparity ($o_{\text{neighbor}} - o_{\text{self}}$) against the resulting opinion change ($\Delta o_{\text{self}}$). 

The baseline scenario shows a relatively high linear correlation ($r = 0.469$), consistent with the direction of assimilative influence \cite{degroot_reaching_1974}. But this aggregate statistic alone is insufficient because it fails to capture the underlying heterogeneity of interaction outcomes. To uncover this heterogeneity, we classify all observations into four quadrants. This allows us to distinguish between two fundamentally different mechanisms: 
(i) \textit{assimilative influence} (first and third quadrants), where agents' opinions move toward their neighbors \cite{degroot_reaching_1974}, and 
(ii) \textit{repulsive influence} (second and fourth quadrants), where agents' opinions move further away \cite{jager_uniformity_2005}.

This quadrant-based decomposition reveals a key departure from traditional models. In bounded-confidence frameworks, interactions typically result in assimilation or no update. In contrast, our simulation generates a substantial share of repulsive responses. As shown in Table~\ref{tab:quadrant_proportions}, even in the baseline scenario (Figure~\ref{fig:Scatter Plot}A), $31.0\%$ of interactions fall into the repulsive quadrants. That is, exposure to opposing views often pushes agents further in the opposite direction. This pattern is consistent with prior findings that repulsive influence is a major driver of macro-level polarization \cite{jager_uniformity_2005}. The dense clusters in the second and fourth quadrants therefore provide a direct micro-level explanation for the polarization observed in aggregate outcomes in Figure \ref{fig:vax_opinion_comparison}.

Beyond the existence of repulsion, the scatter patterns also reveal clear non-linearities. On the right side of the disparity axis (such as in Figure~\ref{fig:Scatter Plot}B, C, and D), moderate opinion differences tend to generate positive opinion updates, indicating the positive impacts brought by the dialogue. However, as disparity increases further, this positive impact breaks down. Extreme differences no longer lead to stronger assimilation; instead, they frequently trigger negative updates. In other words, persuasion exhibits a threshold effect: beyond a certain point, larger disagreement leads to resistance rather than convergence. This pattern is closely aligned with the theory of social judgment \cite{sherif1961}. Moderate differences fall within an agent’s latitude of acceptance and induce assimilation, whereas extreme differences enter the latitude of rejection and provoke defensive resistance. Importantly, this mechanism is not imposed exogenously but emerges naturally from LLM-based interactions.

We further examine how different cognitive modules affect these dynamics. The quadrant statistics in Table~\ref{tab:quadrant_proportions} show that memory module significantly amplifies repulsive responses. The share of repulsion increases from $31.0\%$ in the baseline to $38.6\%$ in the memory scenario (Figure~\ref{fig:Scatter Plot}B) and $38.8\%$ in the combined scenario (Figure~\ref{fig:Scatter Plot}D). At the same time, points in the third quadrant extend further downward, with many approaching the $-1.0$ boundary. This indicates that stronger negative reactions are brought by the dialogue when recalling memories related to vaccination. In contrast, the prompt diversity scenario (Figure~\ref{fig:Scatter Plot}C) reduces repulsion to $29.6\%$. This suggests that heterogeneous communication styles can partially mitigate negative reactions and facilitate more assimilative interactions.

\begin{table}[htbp]
\centering
\caption{Quadrant Proportions Across Scenarios}
\label{tab:quadrant_proportions}
\begin{tabular}{ccccc}
\toprule
\textbf{Scenario} & \textbf{QI+QIII (Assimilative)} & \textbf{QII+QIV (Repulsive)} & \textbf{n} & \textbf{Fig} \\
\midrule
1 & 69.0\% & 31.0\% & 22413 & Fig.8A \\
2 & 61.4\% & 38.6\% & 25259 & Fig.8B \\
3 & 70.4\% & 29.6\% & 22260 & Fig.8C \\
4 & 61.2\% & 38.8\% & 24074 & Fig.8D \\
\bottomrule
\end{tabular}
\end{table}

Finally, the gradual decline in linear correlation (down to $r = 0.341$ in the combined scenario) should not be interpreted as a weakness. Instead, it reflects a shift away from simplified linear assumptions toward richer behavioral details. As agents incorporate memory and diverse communication styles, their responses become less mechanically tied to opinion distance and more behaviorally grounded. Overall, this micro-level evidence demonstrates that our framework not only reproduces classical assimilative dynamics but also captures empirically documented phenomena such as repulsive influence, opinion change thresholds, and resistance to opposing views.


\section{Conclusions and Future Work}
\label{sec:conclusion}

This study introduces an LLM-driven agent-based modeling framework to investigate how different cognitive modules drive macro-level opinion evolution related to vaccination. As shown in Section \ref{sec:results}, our approach leverages large language models to simulate multi-round dialogues and reflections, which replace most of the static numerical rules from traditional agent-based models. Using vaccination opinion evolution as a use case, we successfully demonstrate the framework's capabilities. Specifically, the simulation results from all four scenarios demonstrate the role of different cognitive modules and how each module impacts the overall opinion dynamics, through reporting both macro and micro level results. These micro-results (e.g., influence scatter plots and dialogue logs) also indicate that we could potentially reach the level 3 validation by utilizing LLM to mimic the real dialogue between two people.

Similar to all modeling works, this modeling framework also has room for improvements. One shortage is that the current simulation has only 95 agents, which is a small simulation because the LLM requires extensive computational resources. Thus, one of the future works could be extending the size of the simulation to comprise more synthetic populations (e.g., a city or a county). In addition, there could be opinion leaders in certain circumstances and we did not include this because the aim of this paper is to explore the impact brought by the different cognitive modules. But this can be improved by exploring the structural impact of opinion leaders in future work. Even with these limitations and further works, this study has demonstrated the potentials of integrating LLMs into an agent-based modeling framework by providing an opinion dynamic case study.

\section*{ACKNOWLEDGMENTS}

This work was supported by Guangdong Provincial Talent Program 2025D03J0019. 

\footnotesize

\bibliographystyle{wsc}

\bibliography{demobib}

\section*{AUTHOR BIOGRAPHIES}

\noindent {\bf \MakeUppercase{Bo Zhang}} is a Research Assistant in the Thrust of Urban Governance and Design at The Hong Kong University of Science and Technology (Guangzhou) and Mphil student of Xi'an Jiaotong University. His research focuses on computational social science and agent-based simulation. His e-mail address is \email{bozhang@hkust-gz.edu.cn}.\\

\noindent {\bf \MakeUppercase{Na Jiang}} is an Assistant Professor at the UGOD Thrust of Hong Kong University of Science and Technology (Guangzhou). His research focuses on addressing urban issues utilizing GIS, data science, and computational simulation, particularly agent-based modeling (ABM). His e-mail address is \email{najiang@hkust-gz.edu.cn}.

\end{document}